\documentclass[aps,preprint]{revtex4}

\input{tcilatex}

\begin{document}

\title{Anisotropy and periodicity in the density distribution of electrons
in a quantum-well}
\author{Y. Yayon, M. Rappaport, V. Umansky and I. Bar-Joseph}
\affiliation{Department of Condensed Matter Physics, The Weizmann Institute of Science,
Rehovot 76100, Israel}
\pacs{PACS number}

\begin{abstract}
We use low temperature near-field optical spectroscopy to image the electron
density distribution in the plane of a high mobility GaAs quantum well. We
find that the electrons are not randomly distributed in the plane, but
rather form narrow stripes (width smaller than 150 nm) of higher electron
density. The stripes are oriented along the [1\={1}0 ] crystal direction,
and are arranged in a quasi-periodic structure. We show that elongated
structural mounds, which are intrinsic to molecular beam epitaxy, are
responsible for the creation of this electron density texture.{}
\end{abstract}

\maketitle

\narrowtext

The two-dimensional electron system (2DES) that is formed in a semiconductor
quantum well (QW) is a major platform for studying electron interactions in
two dimensions. The electrons in this structure are confined to a thin
layer, typically 10-20 nm wide, and are free to move only in the direction
parallel to that layer. The significant advance over the last two decades in
growth of semiconductor heterostructures by molecular beam epitaxy (MBE),
particularly of the GaAs/AlGaAs material system, has made a critical
contribution that facilitated these studies. In addition to achieving high
purity and excellent crystalline quality, MBE allows tailoring the potential
and doping profile inside the semiconductor with atomic layer precision.
Using modulation doping the 2DES can be spatially separated from the ionized
impurities, thus strongly enhancing the electron mobility \cite{Stormer}.
Indeed, 2DES samples with high electron mobility, of $10^{6}-10^{7}$ cm$^{2}$%
/(V sec), are readily available today and serve for a variety of studies.

Irregularities in the crystal structure, however, are always present and
introduce a disorder potential even in these high quality samples. The
disorder becomes particularly important and determines the system behavior
when it gives rise to spatial fluctuations in the electron density. A
particular example is the behavior in high magnetic field, which is governed
by the filling factor $\nu $, the ratio between the electron and magnetic
flux densities. Clearly, a spatially varying density would translate into
spatial fluctuations of $\nu $ and lead to inhomogenous behavior.
Nevertheless, despite their importance relatively little is known on the
nature of these electron density fluctuations. It was therefore natural that
with the evolution of scanning-probe experimental techniques a considerable
effort was directed to resolving spatial inhomogeneities in the system
properties \cite{Hess} - \cite{Yacoby}. Near-field spectroscopy \cite%
{Betzig,Paesler} studies have proven to be particularly useful in that
context. The high spatial resolution that can be obtained, typically 100-200
nm, and the wealth of information contained in the optical spectrum made it
a favorable technique for studying the local properties of semiconductor QWs.

In this work we use near-field photoluminescence (PL) spectroscopy to study
spatial correlations in the distribution of the electron density in a gated
2DES. Our ability to extract the electron distribution from the PL spectrum
derives from the fact that a photo-excited electron-hole pair binds to a
``native'' electron in the 2DES and creates a bound complex, known as a
negatively charged exciton, $X^{-}$ \cite{Kheng,Finkelstein}. The $X^{-}$
can be easily identified in the PL spectrum as a peak which appears $\sim 1$
meV below the neutral exciton, $X$, peak (Fig. 1), and its intensity is
directly proportional to the average electron density, $n_{e}$. This
proportionality is demonstrated in the inset of Fig. 1, which shows the area
under the $X^{-}$ peak (measured in the far-field) as a function of $n_{e}$.
We have recently shown that this proportionality between the intensity and
the electron density holds also locally: regions with high electron density
give a large $X^{-}$ signal, and vice versa \cite{Eytan1,Yayon}. This is
demonstrated in Fig. 1 which shows two spectra measured at two nearby points
in the sample. It is seen that the intensity of the $X^{-}$ peak is
significantly different between the two locations while the $X$ peak is
nearly the same. Thus, integrating the area under the $X^{-}$ peak at each
point (shaded area in Fig. 1) yields a value which is proportional to the
local electron density, $I_{X^{-}}(x,y)$. Hence, by photo-exciting the 2DES
with weak laser light and conducting a spatially resolved measurement of the
PL spectrum, we can image the electron distribution in the plane.

The local measurement of the PL spectrum is realized using a homemade
near-field scanning optical microscope (NSOM) \cite{Eytan2} operating at 4.2
K. The sample is illuminated uniformly by a single-mode fiber and the
emitted PL is collected through a tapered Al-coated optical fiber tip, which
is brought into close proximity (\TEXTsymbol{<}10 nm) of the sample surface.
The tip collects the PL and guides it through an optical fiber into a
spectrometer, where it is dispersed onto a liquid nitrogen cooled charged
copule device (CCD) camera. The overall spectral resolution is 80 $\mu $eV.
The spatial resolution is determined by the tip diameter. We have conducted
measurements with tips of various diameters, between 0.2 -- 0.3 $\mu $m. The
two dimensional images shown in this paper were measured with a 0.3 $\mu $m
tip, for which the higher collection efficiency allowed shortening the
acquisition time \cite{Comments-tips}.

The experiment is conducted on a 20 nm GaAs QW. A $\delta $-doped donor
layer with a density of $2.5\times 10^{12}$ cm$^{-2}$ is separated from the
QW by a 50 nm Al$_{0.36}$Ga$_{0.64}$As spacer layer. The distance between
the QW and the sample surface is 96 nm and the total thickness of the MBE
grown layers is 1.55 $\mu $m. A $2\times 2$ $\mu $m$^{2}$ mesa was etched,
and ohmic contacts were alloyed to the 2DES layer. A 4.5nm PdAu
semitransparent gate was evaporated on top of the mesa. By applying a
voltage between the gate and the 2DES we could vary the electron density
under illumination between $1\times 10^{10}$ - $2.5\times 10^{11}$ cm$^{-2}$%
. The electron mobility exhibits anisotropy, being $3\times 10^{6}$ cm$^{2}$%
/(V sec) along the [$1\bar{1}0$] direction and $2\times 10^{6}$ cm$^{2}$/(V
sec) in the orthogonal direction \cite{Comment110}.

Figure 2a shows a two-dimensional image of $I_{X^{-}}(x,y)$, which is
proportional to the electron density, in an area of $11\times 11$ $\mu $m$%
^{2}$. The image shown here was measured at an electron density of $3\times
10^{10}$ cm$^{-2}$. In this low density regime the fluctuations in the
electron density are large compared to the average density, and hence, it is
a convenient regime to measure them \cite{Eytan1}. Similar images have been
measured at different gate voltages, different regions of the same sample,
and on different gated 2DES samples. It is seen that the 2DES density is
highly non-uniform and exhibits large random fluctuations, manifested as
changes of $I_{X^{-}}(x,y)$. The \textit{X} intensity image (not shown), on
the other hand, is very uniform, and shows much smaller and more gradual
fluctuations. This behavior is consistent with the fact the neutral exciton
can diffuse in the plane, and provides a reassuring evidence for the
significance of the observed fluctuations.

To unveil possible order in this seemingly random image we have studied the $%
X^{-}$ intensity autocorrelation function $G(x,y)=<I(x^{\prime },y^{\prime
})I(x^{\prime }-x,y^{\prime }-y)>$, where $<...>$ denotes averaging over all
measured points in the scanned area, and $I(x,y)$ is the difference between $%
I_{X^{-}}(x,y)$ and its intensity average over all points, i.e., $I(x,y)=$ $%
I_{X^{-}}(x,y)-<I_{X^{-}}(x,y)>$. The correlation function averages out the
random behavior and highlights the correlated part of the signal. The width
of its central peak provides the correlation length, namely, the typical
cluster size, and periodic peaks in $G(x,y)$ indicate the existence of
periodicity in the electron distribution. The function $G(x,y)$\ is shown in
Fig. 2b, with bright and dark colors signifying high and low correlation
amplitude, respectively. It is seen that there is a pronounced anisotropy in 
$G(x,y)$: it is dominated by periodic stripes and its central peak is
elongated parallel to [$1\bar{1}0$]. To better understand this behavior we
show in Fig. 2c two cuts through the origin of $G(x,y)$, one along [$1\bar{1}%
0$] (dashed) and the other perpendicular to it, along [$110$] (solid). We
can see that both the short and long range correlations are different. The
cut along the [$110$] direction exhibits a correlation length of 0.35 $\mu $%
m (half-width at half maximum), and clear long-range oscillations with a
periodicity of $\sim 1.3$ $\mu $m. The [$1\bar{1}0$] cut, on the other hand,
shows a significantly larger correlation length, 0.6 $\mu $m, and no
significant long-range order. We have verified that this behavior is not an
instrument artifact by both rastering at different angles and by rotating
the sample relative to the scan direction. Furthermore, no stripes were
observed in the $X$ intensity image. The correlation length along the [$110$%
] direction, 0.35 $\mu $m, is only slightly larger than the tip diameter,
which is 0.3 $\mu $m. In the inset of Fig. 2c we compare the central peak
measured with tips having 0.2 and 0.3 $\mu $m diameters. The correlation
length is seen to scale with the tip diameter, and it is 0.25 $\mu $m with
the smaller tip. This allows us to determine the width of the stripes, after
deconvolving the tip diameter, to be smaller than 150 nm.

The formation of electron density stripes is further emphasized by making
cuts in $G(x,y)$ at different angles, $\theta $. This is formally defined as
the angular autocorrelation function $H(r,\theta )=<I(x^{\prime },y^{\prime
})I(x^{\prime }-r\cos \theta ,y^{\prime }-r\sin \theta )>$. The results are
shown in Fig. 2d, with the horizontal scale being $r$ and the vertical scale
- $\theta $. Clear concentric contours are observed, describing a gradual
change of the periodicity with $\theta $. It is easy to see that this
behavior indeed corresponds to periodic electron density stripes along the [$%
1\bar{1}0$] direction. Clearly, the oscillation period of such a stripes
pattern would vary with $\theta $ as $\lambda /\cos \theta $, where $\lambda 
$ is the period of the stripes.

The fact that the electron density stripes coincide with crystalline
orientation and the relatively large periodicity point to an underlying
crystalline structure as the source of this behavior. To examine this
possibility we performed atomic force microscopy (AFM) measurements of the
crystal surface on different pieces of the same wafer. Figure 3a shows a
typical topography image and Fig. 3b -- the variations in cantilever
amplitude, which are proportional to the topography derivative. It is seen
that narrow ridges, which are oriented along the [$1\bar{1}0$] direction,
cover the sample surface. The formation of these ridges is well studied and
understood. Their origin is an intrinsic growth instability, which inhibits
downward movement of adatoms at surface step edges, and leads to build-up of
mounds \cite{Johnson}. The heights of these mounds and their typical size
increase with the number of layers grown \cite{Orme}, and their anisotropy
is due to the particular surface reconstruction of the As atoms that
terminate the (001) surface \cite{Comment-arsenic}. The dangling bonds of
these atoms form dimers that are arranged in rows along the [$1\bar{1}0$]
direction, and since the growth rate along the dimer rows is larger than in
the orthogonal direction, the mounds become elongated along [$1\bar{1}0$].
Clearly, these mounds modulate the distances of the ionized donors layer and
the crystal surface relative to the 2DES plane, and thereby cause variations
in the electrostatic potential in that plane. Considering the structure as a
parallel plate capacitor (gate and 2DES) with a positively charged layer
(the donors) in between, it is straightforward to show that varying the
distances of these layers in our sample by a monolayer would be translated
to a density change of $\sim 10^{9}$ cm$^{-2}$ in the 2DES. The relation
between the density distribution and the crystal topography is best
demonstrated by comparing the autocorrelation functions of the density
distribution and that of the topography image. Fig. 3c shows two cuts of the
topography autocorrelation function along [$1\bar{1}0$] (dashed) and [$110$]
(solid). The similarity between Figs. 2c and 3c is evident. We notice,
however, that the periodicity in the topography image is considerably larger
than that of the electron density stripes. This larger period was found in
all AFM measurements of different areas of the same wafer. The difference
between the surface mounds and the electron density periods is not
surprising. The coarsening process that occurs during the growth, in which
small mounds join to form larger ones, results in an increase of the period
towards the surface. Hence, the distance between the 2DES and the surface
would be modulated with a period, which is roughly the average between the
mounds period at the surface and at the 2DES.

It should be emphasized that other disorder potentials might also affect the
electron distribution. Fluctuations in the QW width cause the electron
confinement energy in the well to vary from one point to another and hence
introduce potential fluctuations. These fluctuations can be imaged as well
in our experiment by determining the \textit{X} peak energy at each point %
\cite{Hess,Wu}. We have found that the well width fluctuations, despite
being sensitive to the mounds \cite{Comment well width}, show no correlation
with the charge fluctuations. The QW width fluctuations as well as the
fluctuations in the donor density are the probable source of the random
density pattern, which is evident in Fig. 2a but averages out in the
correlation function.

It is interesting to consider our results in the context of the recent
debate on the origin of transport anisotropy in high Landau levels \cite%
{Lilly1}-\cite{Willet}. Our observation of charge stripes provides a direct
support to the suggestion that the transport anisotropy is due to the
underlying crystalline structure 20. Finally, we wish to note that the
elongated mounds are intrinsic property of MBE growth and are found even at
the highest mobility samples \cite{Lilly2,Willet}. Hence, the electron
density stripes formation is a fundamental property of 2DES systems, and
should play a particularly important role at low electron densities, where
the electrons form a network of quasi-one dimensional wires.

We are pleased to acknowledge D. Kandel for helpful discussions on the MBE
growth, E. Khivrich for his assistance in the AFM measurements, and G. Fish
from Nanonics for providing the NSOM tips.


\begin{thebibliography}{99}
\bibitem{Comments-tips} The tip size was determined by measuring the PL from
local defects.
\bibitem{Comment110} The determination of the [1\={1}0] direction was done
by identifying the ``bowtie'' defects on the back of the sample.
\bibitem{Comment-arsenic} The arsenic overpressure causes the crystal
surface to be terminated by arsenic atoms.
\bibitem{Comment well width} The behavior of the well width fluctuations
will be reported elsewhere.
\end{thebibliography}
\end{document}